\documentclass[aps,pra,twocolumn,superscriptaddress,showpacs,letterpaper]{revtex4}

\usepackage{amsmath}
\usepackage{color,graphicx}
\usepackage[pdfstartview=FitH,colorlinks=true,linkcolor=blue,citecolor=blue,urlcolor=blue]{hyperref}

\begin{document}

\title{Dark-Bright Solitons in a Superfluid Bose-Fermi Mixture}

\author{Marek Tylutki}
\email{marek.tylutki@ino.it}
\affiliation{INO-CNR BEC Center and Dipartimento di Fisica, Universit\`a di Trento, Via Sommarive 14, I-38123 Povo, Italy}

\author{Alessio Recati}
\affiliation{INO-CNR BEC Center and Dipartimento di Fisica, Universit\`a di Trento, Via Sommarive 14, I-38123 Povo, Italy}
\affiliation{Technische Universit\"at M\"unchen, James-Franck-Stra{\ss}e 1, 85748 Garching, Germany}

\author{Franco Dalfovo}
\affiliation{INO-CNR BEC Center and Dipartimento di Fisica, Universit\`a di Trento, Via Sommarive 14, I-38123 Povo, Italy}

\author{Sandro Stringari}
\affiliation{INO-CNR BEC Center and Dipartimento di Fisica, Universit\`a di Trento, Via Sommarive 14, I-38123 Povo, Italy}

\date{\today}

\begin{abstract}
The recent experimental realization of Bose-Fermi superfluid mixtures of dilute ultracold atomic gases has opened new perspectives in the study of quantum many-body systems. Depending on the values of the scattering lengths and the amount of bosons and fermions, a uniform Bose-Fermi mixture is predicted to exhibit a fully mixed phase, a fully separated phase or, in addition, a purely fermionic phase coexisting with a mixed phase. The occurrence of this intermediate configuration has interesting consequences when the system is nonuniform. In this work we theoretically investigate the case of  solitonic solutions of coupled  Bogoliubov-de Gennes and Gross-Pitaevskii equations for the fermionic and bosonic components, respectively. We show that, in the partially separated phase, a dark soliton in Fermi superfluid is accompanied by a broad bosonic component in the soliton, forming a dark-bright soliton which keeps full spatial coherence.  
\end{abstract}


\maketitle


\section{Introduction}

A long standing problem in the context of quantum fluids is the description of mixtures composed of two kinds of interacting superfluids belonging to different statistics. The first theoretical analysis of superfluid $^4$He mixed with superfluid $^3$He dates back to the 70's (see, for example, \cite{AndreevBashkin,Khalatnikov} and references therein). In experiments, however, the simultaneous superfluidity of the two components of liquid $^3$He-$^4$He mixtures has never been realized, since the miscibility of $^3$He in  $^4$He is very small (a few percent) and the temperature needed to reach superfluidity of fermions in the mixture is too low to be reached with available cryogenic techniques.  Dilute ultracold atomic gases are instead excellent candidates for studying superfluid properties of mixtures. Superfluidity has been recently obtained experimentally in a mixture of a Bose condensed gas and a  superfluid Fermi gas of two lithium isotopes, $^6$Li and $^7$Li \cite{Ferrier-Barbut2014,Delehaye2015}, where a new mechanism for superfluid instability was observed, related to the dynamical instability of the supercurrent counterflow rather than to the more standard Landau criterion~\cite{Castin2015,Marta_CF}. In addition, in ultracold atomic gases the strength of the interspecies and intraspecies interaction can be varied by means of an external magnetic field, thanks to the occurrence of Feshbach resonances. One can thus foresee the exploration of the whole phase diagram of the mixture, which is expected to be very rich~\cite{Viverit2000,Salasnich2007,SalasnichBoseFermiSF,QCDBaym,Ludwig2011,PuSFBoseFermi,Bertaina2013,Ozawa2014,Akdeniz2005}.  

In the present work, we focus on the case where the bosonic superfluid is the minority component, while the fermionic superfluid exhibits a dark soliton. We choose this case as a paradigmatic configuration in which the interplay between miscibility and immiscibility, together with superfluidity, gives rise to a peculiar behaviour which reveals the crucial effects of nonlinearity caused by interactions. The opposite limit of bright solitons in a mixture with fermions as the minority component was discussed in~\cite{Karpiuk2004}.

In pure Fermi superfluids of dilute atomic gases, theoretical predictions of the structure and dynamics of dark solitons~\cite{Dziarmaga2005,Antezza2007,Sacha2010,Scott2011,Spuntarelli2011,Liao2011,Scott2012,Scott2013}  have recently stimulated experimental investigations~\cite{Yefsah2013}. The experiments confirm the theoretical expectation that dark solitons in a three-dimensional fermionic superfluid quickly decay into vortical excitations due to snaking instability~\cite{Ku2015}, as it was earlier observed with bosons~\cite{Dutton2001,Anderson2001}. With bosons very long-lived dark solitons have been generated by filling the soliton with atoms in another hyperfine state~\cite{Becker2008,Middelkamp2011,Hamner2011}, thus creating a dark-bright solitonic structure of a two-component Bose-Bose superfluid~\cite{Busch2001}. Here we theoretically investigate the analogue dark-bright soliton in a Bose-Fermi superfluid mixture, the main difference between the two cases being that the Bose-Fermi phase diagram is known to admit, in addition to a fully mixed phase and a fully separated phase, also a third phase consisting of pure fermions in equilibrium with a mixture of fermions and bosons. The stability conditions of such an intermediate phase in a uniform system were studied in \cite{Viverit2000} by using the equation of state of an ideal Fermi gas weakly interacting with a dilute Bose gas. Such a phase is predicted to occur also in the strongly interacting regime~\cite{Ludwig2011}. A more refined equation of state, including the interaction among fermions, was later applied also to nonuniform configurations by treating the interaction energy in local density approximation~\cite{Salasnich2007}. However, since a dark soliton is localized on the length scale of the healing length of the superfluid, which is of the order of the inverse Fermi wave vector for fermions at unitarity, its characterization requires a theory which properly includes non-local effects, beyond the local density approximation. For this purpose, we use coupled Bogoliubov-de Gennes and Gross-Pitaevskii equations for the fermionic and bosonic components respectively. 

Our paper is organized as follows: in \S~\ref{sec.EoS} we discuss the stability condition of the uniform phase of a Bose-Fermi mixture when the Fermi gas is at unitarity (infinite scattering length); in \S~\ref{sec.mf} we write the mean-field equations, which are subsequently used to find the stationary solitonic configurations of the system; in \S~\ref{sec.unitarity} we analyse the behaviour of solitons of a Bose-Fermi mixture when fermions are at unitarity. We pay special attention to the transition from the miscible state to the so called partially separated phase and to the fully separated state. We find that, in the partially separated phase, the density depletion of fermions possesses a solitonic character and the phase coherence between the left and right hand sides is maintained; conversely, in the fully separated phase, the depletion in the Fermi density is completely filled by bosons (the density of the fermions vanishes) and the phase coherence between the two sides is lost. Finally in \S~\ref{sec.bcsbec} we investigate the behaviour of solitons along the crossover from the Bardeen-Cooper-Schrieffer phase to Bose-Einstein condensation (BCS-BEC crossover), and we conclude that, while on the BCS side the system exhibits a behaviour similar to that at unitarity, on the BEC side the partially separated configuration disappears as expected for a Bose-Bose mixture.  


\section{Phase Separation}
\label{sec.EoS}
Let us first discuss the conditions for the stability of the homogeneous phase of the mixture, which is a crucial point in order to understand the  numerical results of the following sections. The phase diagram of a weakly interacting Bose-Fermi mixture, at zero temperature, can be derived starting from the following expression for the energy density:
\begin{equation}
\mathcal{E}[n_f, n_b] = \frac12 g_{bb} n_b^2 + g_{bf} n_b n_f + \frac{3}{5} \eta E_F n_f ~.
\label{eq.eos-bf}
\end{equation}
Here $n_b$ and $n_f$ are the densities of the Bose and Fermi gas respectively, and $E_F=\hbar^2 k_F^2 /2m_f$ is the Fermi energy, with $k_F=(3\pi^2 n_f)^{1/3}$. In the above expression we assume that the Fermi gas is at unitarity (infinite scattering length) and $\eta$ is the dimensionless Bertsch parameter~\cite{Carlson2003,Astrakharchik2004,Giorgini2008}, which simply rescales the energy density of the Fermi gas with respect to the ideal gas expression. The quantities $g_{bb}$ and $g_{bf}$ are the bosonic intraspecies and the Bose-Fermi interspecies coupling constants, which are related to the corresponding scattering lengths  according to $g_{bb} = 4 \pi \hbar^2 a_{bb} / m_b$ and $g_{bf} = 4 \pi \hbar^2 a_{bf} (m_b + m_f)/ (2 m_b m_f)$, where $m_f$ and $m_b$ are the masses of fermions and bosons. We assume that the Bose-Fermi scattering length does not depend on the internal state of the Fermi atoms, as in the case of the recent experiments with lithium atoms~\cite{Ferrier-Barbut2014,Delehaye2015}. Equation (\ref{eq.eos-bf}) was first used in~\cite{Viverit2000} and ~\cite{Molmer1998} to describe the phase diagram of a dilute Bose gas interacting with an ideal Fermi gas ($\eta = 1$). In~\cite{Viverit2000} the phase diagram was explored as a function of the densities of the two components and the existence of three phases was predicted for a positive Bose-Fermi scattering length: i) a uniform mixture, where both components occupy the entire space at constant densities; ii) a partially separated phase, where part of the space is occupied by pure fermions and part by a Bose-Fermi mixture; iii) a fully separated phase, where bosons and fermions are completely separated. The same happens at unitarity, with the only difference that the Fermi energy is renormalized by the universal Bertsch parameter $\eta$. It is worth noticing that the existence of three phases is peculiar of the Bose-Fermi mixture. In fact a Bose-Bose mixture only admits the mixed uniform phase and the fully separated phase, because of the different power-law dependence on the densities in the equation of state. 

The stability condition predicted by the energy density~(\ref{eq.eos-bf}) for the uniform mixture is 
\begin{equation}
n_f^{1/3} \leq \frac{2}{3}\, (6 \pi^2)^{2/3}\, \eta\, \frac{\hbar^2}{2m_f}\,  \frac{g_{bb}}{g_{bf}^2} ~.
\label{eq.lda-bertsch}
\end{equation}
For $n_f$ larger than this critical value the uniform mixture is unstable and the system exhibits either  partial or full phase separation. 

\begin{figure*}[!t]
\includegraphics[width=.95\textwidth]{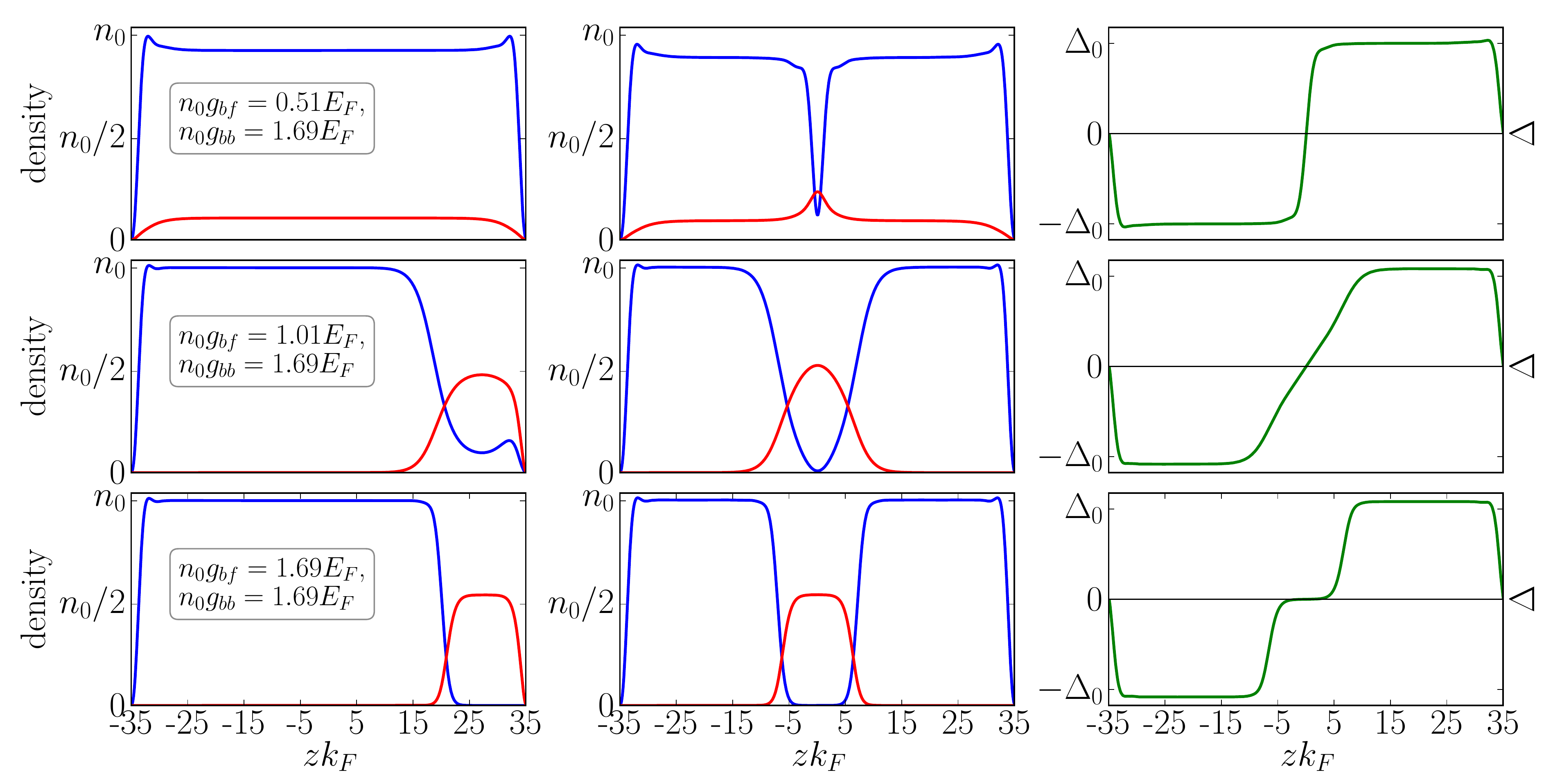}
\caption{(color online) Left column: Density profiles of the ground state of fermions (blue line) and bosons (red line), with fermions at unitarity ($1/k_F a_{ff} = 0$), for three sets of interaction parameters such that $n_0 g_{bf}^2 / (g_{bb} E_F)=0.15$, $0.6$ and $1.69$, from top to bottom. The number of bosons is $10 \%$ of the number of fermions. The three panels represent examples of fully mixed, partially mixed and fully separated phases. Central columns: Same as before, but for stationary states with a dark soliton in the Fermi component. In the top panel the dark soliton is slightly modified by bosons; in the central panel the two superfluids form a dark-bright soliton; in the bottom panel the two components are fully separated. Right column: order parameter $\Delta$ of the Fermi superfluid for the same configurations of the central column; the value in the bulk of pure fermions is $\Delta_0 = 0.68 E_F$.}
\label{fig.unitarity}
\end{figure*}

\begin{figure}
\includegraphics[width=.95\columnwidth]{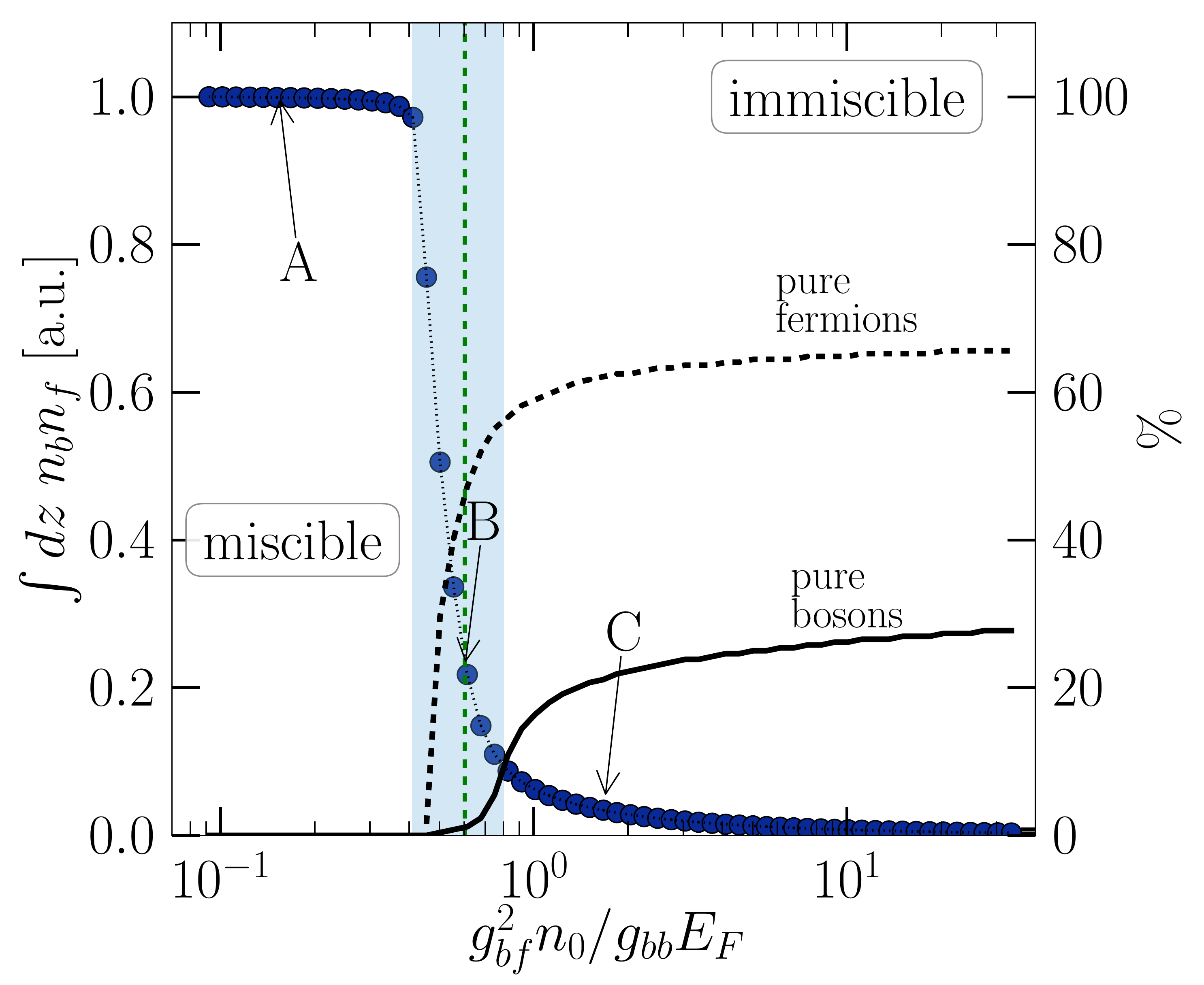}
\caption{(color online) Transition from miscible to immiscible phases of the mixture. The overlap integral $\int d z\, n_f(z) n_b(z)$ is calculated in the ground state obtained for different values of $g_{bf}^2 n_0 / (g_{bb} E_F)$, and the results are shown as markers (scale on the left axis). When this quantity deviates from $1$ the fully mixed phase becomes unstable. The vertical green dashed line indicates the instability threshold predicted by Eq.~(\ref{eq.lda-bertsch}) in the thermodynamic limit, with $\eta=0.59$. For large values of $g_{bf}^2 n_0 / (g_{bb} E_F)$ the two components fully separate and the overlap eventually vanishes. The three letters A, B and C indicate the solutions reported in the left column of Fig.~\ref{fig.unitarity}. The thick dashed and solid lines show the percentage of volume occupied by pure fermions and pure bosons, respectively (scale on the right axis). Near the instability threshold, a finite interval of $g_{bf}^2 n_0 / (g_{bb} E_F)$ exists where a significant part of the box is filled by pure fermions, while bosons are still in a mixed phase in the remaining volume; this corresponds to the partially mixed phase of the mixture. The vertical shaded area is the region where we find a stable dark-bright soliton.}
\label{fig.overlap-unitarity}
\end{figure}


\section{Mean-field equations}
\label{sec.mf}

In order to describe both uniform and nonuniform configurations of the mixture, including solitons,  we use the Bogoliubov-de Gennes (BdG) equations~\cite{Giorgini2008} for the interacting superfluid fermions and the Gross-Pitaevskii (GP) equation~\cite{Dalfovo1999} for the Bose-condensed bosons, which are coupled {\it via} the interspecies interaction term, fixed by $g_{bf}$. We also use a simple geometry consisting of a three-dimensional rectangular box with hard walls in the longitudinal direction $z$ (box size $L$) and periodic boundary conditions in the transverse directions $x$ and $y$ (box size $L_\perp$). The system is  assumed to be uniform in the transverse directions and all spatial variations are along $z$.

The Fermi gas is described in terms of a set of quasiparticle amplitudes $\{u_j\}$ and $\{v_j\}$, solutions of the BdG equations, which in our case take the form: $u_{j,{\bf k}_\perp}({\bf r}) = (\sqrt{k_F}/L_\perp) e^{i {\bf k_\perp}\cdot {\bf r_\perp}} u_{j,{\bf k}_\perp}(z)$ and $v_{j,{\bf k}_\perp}({\bf r}) = (\sqrt{k_F}/L_\perp) e^{i {\bf k_\perp}\cdot {\bf r_\perp}} v_{j,{\bf k}_\perp}(z)$, with ${\bf r} = ({\bf r_\perp}, z)$ and $j$ is the generic quantum number associated to the longitudinal degrees of freedom. With this choice, the BdG equations become 
\begin{subequations}
\begin{equation} 
\begin{bmatrix}
H & \Delta\\
\Delta^* & -H \\
\end{bmatrix} \begin{bmatrix}
u_{j,{\bf k}_\perp}\\
v_{j,{\bf k}_\perp}\\
\end{bmatrix} = \varepsilon_{j,{\bf k}_\perp} \begin{bmatrix}
u_{j,{\bf k}_\perp}\\
v_{j,{\bf k}_\perp}\\
\end{bmatrix}
\label{eq.bdg}
\end{equation}
where
$H = \hbar^2 ({\bf k}_\perp^2 - \partial_z^2) / 2 m_f - \mu_f + g_{bf} |\psi_b|^2$ is an effective single quasiparticle grand-canonical Hamiltonian, while  
\begin{eqnarray} 
\Delta(z) &=& -\frac{g_{ff}}{L_\perp^2} \sum_{j,{\bf k}_\perp} u_{j,{\bf k}_\perp}(z) \, 
v^*_{j,{\bf k}_\perp}(z) ~,\label{eq.delta}\\
n_f(z) &=& \frac{2}{L_\perp^2} \sum_{j,{\bf k}_\perp} |v_{j,{\bf k}_\perp}(z)|^2 ~,\label{eq.dens}\\  \nonumber
\end{eqnarray}
are the order parameter (gap function) and the density of the superfluid fermionic component. The quantity $\mu_f$ is the chemical potential of fermions, and the eigenvalues $\varepsilon_{j,{\bf k}_\perp}$ are the quasiparticle energies.  

The Bose-condensed component is instead described by the Gross-Pitaevskii equation for the bosonic order parameter (macroscopic wave function) $\psi_b$:
\begin{equation} 
-\frac{\hbar^2}{2m_b} \nabla^2 \psi_b + g_{bb} |\psi_b|^2 \psi_b + g_{bf} n_f \psi_b = \mu_b \psi_b \\ ,
\label{eq.gpe} 
\end{equation}
\end{subequations}
where $\mu_b$ is the chemical potential of bosons and the density is $n_b(z)=|\psi(z)|^2$.  

The four equations~(\ref{eq.bdg})-(\ref{eq.gpe}) must be solved self-consistently. An energy cutoff $E_{\rm c} = 50 E_F$ is used to solve the BdG equations; this implies a proper renormalization of the Fermi-Fermi coupling constant, for which we use the relation $1 / (k_F a_{ff}) = 8 \pi E_F / (g_{ff} k_F^3) + 2 \sqrt{E_{\rm c} / E_F} / \pi$ \cite{Antezza2007}. The key parameter characterizing the interaction among fermions is $1 / {k_F a_{ff}}$, and we perform calculations in the range $-1 \leq 1 / {k_F a_{ff}} \leq 1$, in the crossover from the BCS regime (negative values) to the BEC regime (positive values), passing through unitarity ($1 / {k_F a_{ff}} =0$). 


\section{Solitonic solutions in the unitarity regime}
\label{sec.unitarity}

In the following we will consider a mixture where the number of bosons is about $10 \%$ of the number of fermions. For simplicity we also impose $m_b=m_f$, which is a reasonable approximation for mixtures of two isotopes of the same atomic species, but this assumption does not affect the main results of the work. We typically solve Eqs~(\ref{eq.bdg})-(\ref{eq.gpe}) with $N_f \approx 500$ fermions and $N_b \approx 50$ bosons in a box with $L = 70 k_F^{-1}$ and $L_\perp = 15 k_F^{-1}$. We start from trial functions $\Delta(z)$ and $\psi_b(z)$ and iterate till convergence to a stationary solution which does not depend on the initial choice. 

Examples are given in Fig.~\ref{fig.unitarity} for the case of fermions at unitarity. The Fermi wave vector $k_F$ is the same in all figures and is related to the bulk density of fermions in the pure phase, $n_0$,  by $k_F^3 = 3 \pi^2 n_0$. The corresponding bulk value of the order parameter at unitarity is $\Delta_0 = 0.68 E_F$. Let us concentrate on the first column where we plot the density profiles of fermions (blue lines) and bosons (red lines) for the ground state in the box, for three different values of the parameter $g_{bf}^2 n_0 / (g_{bb} E_F)$. The three values are chosen as representative of the different phases: a fully mixed phase for the smallest value of $g_{bf}^2 n_0 / (g_{bb} E_F)$ (top panel), a fully separated phase for the largest one (bottom panel), and an intermediate partially separated phase (middle panel). In the latter case fermions occupy the whole volume, partly as a pure phase and partly in a mixed phase, and the interface between the two regions is significantly wider than the domain wall found for the fully separated phase (bottom panel), where the width of the interface is of the order of the healing lengths of the two superfluids. The healing length of bosons is much smaller than the size of the box in our case, because we have chosen the value of $g_{bb}$ such that the solution of the GP equation is well approximated by the Thomas-Fermi approximation $n_b(z) = \left( \mu_b - g_{bf} n_f(z) \right) / g_{bb}$, except near the box boundaries. 

To gain further insight, we perform calculations for several values of $g_{bf}^2 n_0 / (g_{bb} E_F)$ and, in each case, we calculate the overlap integral $\int dz\, n_f (z)n_b(z)$. The results are shown as markers in Fig.~\ref{fig.overlap-unitarity}. The overlap integral is $1$ in the fully mixed phase and vanishes in the fully separated phase. The instability of the uniform mixture is clearly visible as a sharp deviation from $1$, which occurs at the critical value of $g_{bf}^2 n_0 / (g_{bb} E_F) \simeq 0.4$. In an infinite system (i.e., in the thermodynamic limit, where surface and interface effects are ignored) this value is expected to be well approximated by the linear stability condition (\ref{eq.lda-bertsch}); by using the mean-field value of the Bertsch parameter, $\eta=0.59$~\cite{Giorgini2008}, this threshold is $g_{bf}^2 n_0 / (g_{bb} E_F) \simeq 0.6$ (vertical green dashed line). In the same figure we also show the percentage of space occupied by the pure Fermi gas (thick dashed line) and the pure Bose gas (solid line). One can see that, by increasing $g_{bf}^2 n_0 / (g_{bb} E_F)$ above the critical value $\simeq 0.4$, pure fermions start occupying a significant part of the box while bosons remain still mixed with fermions, which is another indication of the occurrence of the intermediate phase. 

Having discussed the ground state as a test case, let us now concentrate on the second column of Fig.~\ref{fig.unitarity}, where we show the density profiles in the presence of a dark soliton in the Fermi component. The parameters are the same as in the first column. The dark soliton is imprinted in the Fermi superfluid by imposing a node ($\Delta = 0$) of the order parameter at the centre of the box ($z = 0$) and a $\pi$ phase difference between the two sides. In a purely fermionic superfluid, the stationary solution of the BdG equation with such constraints exhibits a deep density depletion of width of the order of $k_F^{-1}$~\cite{Antezza2007}. If bosons are present, depending on the values of the interaction strengths $g_{bf}$ and $g_{bb}$, they can be attracted into the soliton, thus changing its structure. The figure shows that, for small values of $g_{bf}$ (top panel), the system favours a uniform mixed configuration in agreement with the condition (\ref{eq.lda-bertsch}) and only a small fraction of bosons is attracted by the soliton, which remains almost unaltered. The resulting structure is analog to the dark--anti-dark soliton pairs that are predicted for Bose-Bose mixtures~\cite{Kevrekidis2003,Kasamatsu2006}. In the opposite limit of large $g_{bf}$ (bottom panel) all bosons form a pure phase at the centre, pushing fermions apart, and the soliton is replaced by two domain walls separating pure bosons from pure fermions. The central case is the most interesting: bosons like to stay mixed with fermions, but fermions like to form a pure phase. The net effect is that all bosons are pushed into the soliton but with a broad overlap between the two components. The overall structure keeps its solitonic character and becomes a bright-dark soliton. The shaded area in Fig.~\ref{fig.overlap-unitarity} is the interval of $g_{bf}^2 n_0 / (g_{bb} E_F)$ where we find stationary solutions having such a bright-dark soliton structure, with both the node of the order parameter and the minimum of the fermionic density at a single point, $z=0$. The resulting scenario shares interesting analogies with the dark-bright solitonic structure exhibited by two-component Bose superfluids~\cite{Busch2001}. We find similar solutions also for different values of the bosonic fraction: $N_b = 20 \%$, $30 \%$ and $50 \%$ of $N_f$. 

For the three dark soliton solutions in the second column of Fig.~\ref{fig.unitarity} the phase of the order parameter is $\pi$ for negative $z$ and $0$ for positive $z$, which implies that $\Delta(z)$ is a real function. The third column of the same figure shows the corresponding profiles of $\Delta(z)$. However, in the case of complete separation (bottom panel) the phase difference between the two sides is irrelevant, because the two regions of pure fermions, separated by the central Bose gas, behave as independent superfluids with no phase coherence; in fact, we have numerically checked that, by arbitrarily changing the phase difference of the order parameter, both $n_f$ and $|\Delta|$ remain unchanged in the solution of the coupled BdG and GP equations. Moreover, adding more bosons would simply result in further separating the fermionic superfluids, keeping the same structure of the domain walls. This is not the case of the partially mixed phase shown in the central panel where, if we suddenly change the sign of $\Delta$ on the left side, for instance, the solitonic structure is quickly lost and the solution converges to one without soliton, as in the first column. This proves that the dark-bright soliton in the intermediate partially mixed phase has indeed solitonic character, being an effect of nonlinearity and phase coherence. 

\begin{figure}
\includegraphics[width=.9\columnwidth]{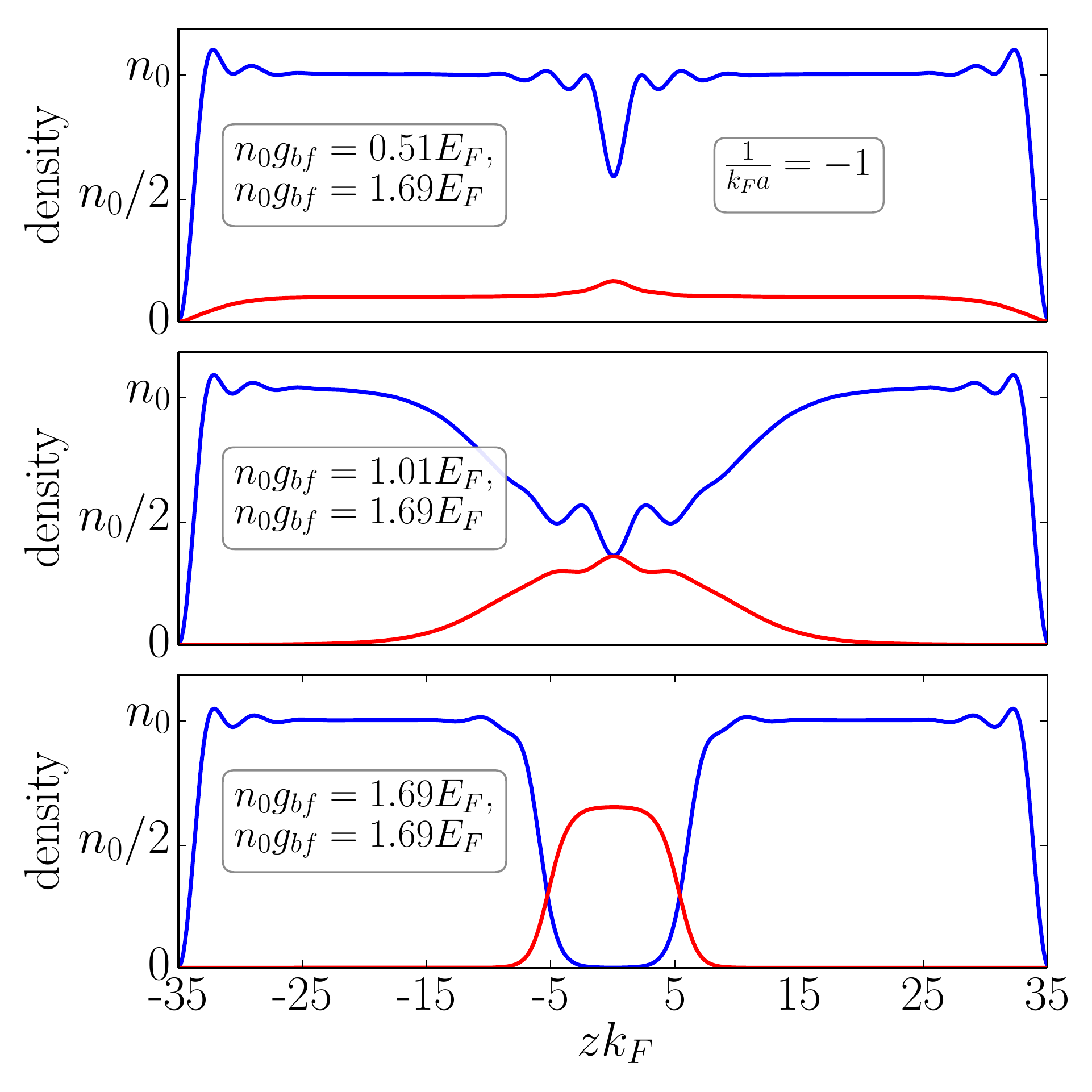}
\caption{(color online) Density profiles of fermions (blue line) and bosons (red line) for solitonic states as in the central column of Fig.~\ref{fig.unitarity}, with the same interaction parameters, but with fermions in the BCS regime ($1/k_F a_{ff} = -1$). The structure is qualitatively the same as at unitarity, but the soliton is shallower and the fermionic density exhibits Friedel oscillations. The order parameter, not shown, is also similar to the one in the right column of Fig.~\ref{fig.unitarity}, but with $\Delta_0 = 0.22 E_F$.}
\label{fig.bcs}
\end{figure}


\section{Fermions in the BCS and BEC regimes}
\label{sec.bcsbec}

Now we discuss what happens when the fermionic component of the mixture is in the BCS-BEC crossover, away from unitarity. 

In the BCS regime ($1 / {k_F a_{ff}} < 0$) fermions can also phase-separate from bosons either partially or completely. This is not surprising, as the BCS and the unitary regimes share many similar features as suggested, for example, by the fact that the chemical potential is positive in both cases. The qualitative picture of phase separation is the same: a uniform mixture is stable for small values of the the parameter $g_{bf}^2 n_0 / (g_{bb} E_F)$, a partial phase separation occurs for larger values, and a complete phase separation for an even larger coupling. As a consequence, also the structure of the solitonic solutions is expected to be similar. In Fig.~\ref{fig.bcs} we show the results for $1 / {k_F a_{ff}} = -1$, for the same parameters used at unitarity in the central column of Fig.~\ref{fig.unitarity}. Apart from more pronounced Friedel oscillations and a shallower solitonic depletion in the density distribution of fermions~\cite{Antezza2007}, we find that the results look indeed very similar. If one moves further into the BCS regime, however, the solitonic structure becomes broader and broader, and less robust against instability mechanisms associated to the fermionic degrees of freedom~\cite{Scott2012}.     

\begin{figure}
\includegraphics[width=.9\columnwidth]{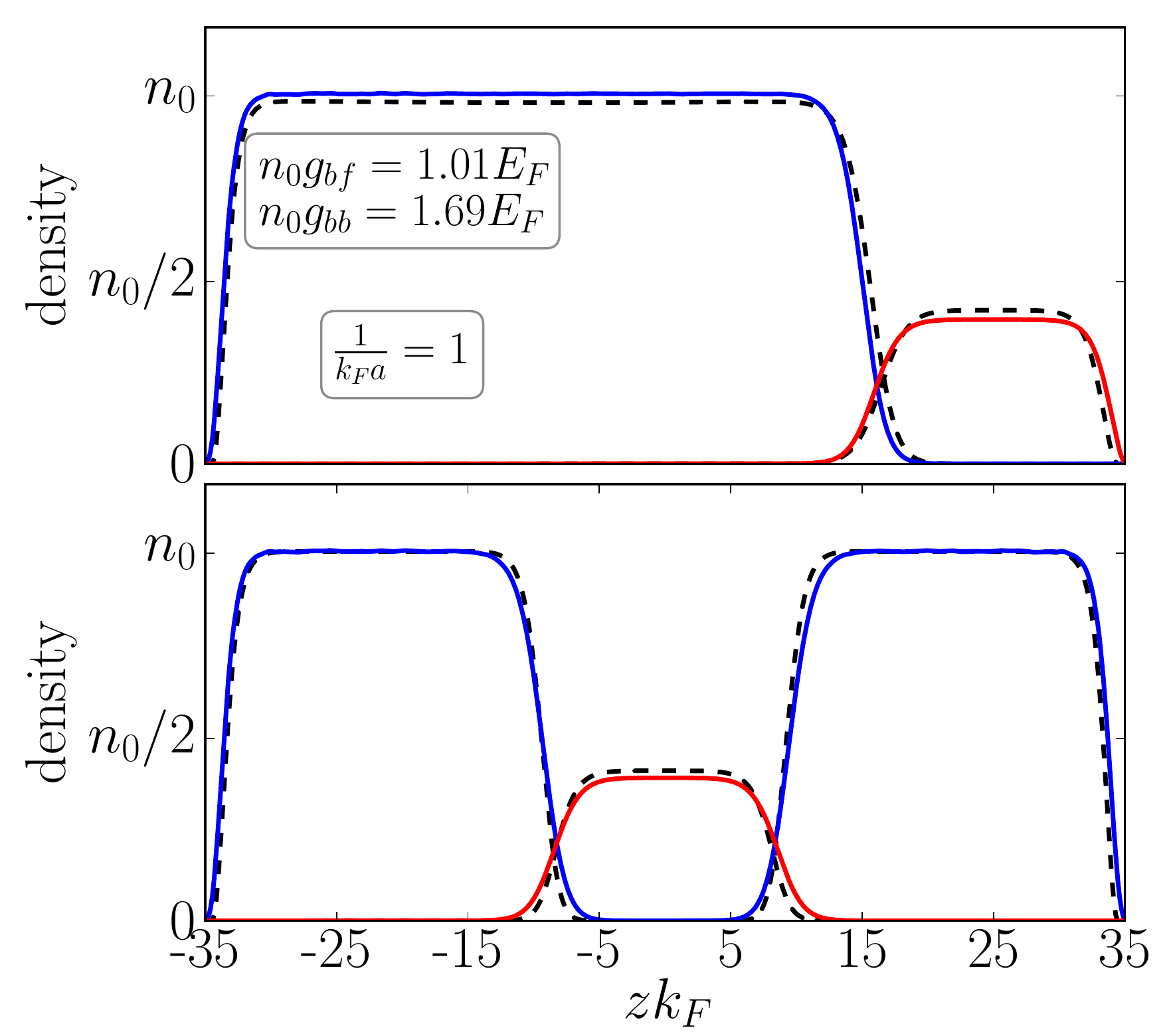}
\caption{(color online) Density profiles of fermions (blue line) and bosons (red line) for uniform (top) and solitonic (bottom) states as in the central row of Fig.~\ref{fig.unitarity}, with the same interaction parameters, but with fermions in the BEC regime ($1/k_F a_{ff} = 1$). The black dashed lines show the results obtained with the coupled GP equations~(\ref{eq.2gpe}).}
\label{fig.bec}
\end{figure}

In order to discuss the BEC regime ($1 / {k_F a_{ff}} > 0$) we have to recall that the fermionic superfluid of ultracold atoms in the BCS-BEC crossover is actually made of an equally populated two-spin-component Fermi gas and the $s$-wave scattering length $a_{ff}$ accounts for the interaction between atoms with different spins. Due to pairing, this two-component gas behaves as superfluid described by an order parameter $\Delta$ and the total density $n_f$, as we have done so far. However, in the BEC regime fermions with opposite spins form tightly bound bosonic dimers, which in turn form a Bose-Einstein condensate. In this situation neither Eq.~(\ref{eq.eos-bf}) nor Eq.~(\ref{eq.lda-bertsch}) hold. The Bose-Fermi mixture thus behaves as a Bose-Bose mixture, where one of the two bosonic components is the condensate of dimers, and the system can be described by two coupled GP equations 
\begin{eqnarray}\nonumber
i \hbar \partial_t \psi_d &=& - \frac{\hbar^2}{2m_d} \nabla^2 \psi_d + g_{dd} |\psi_d|^2 \psi_d + \tilde{g} |\psi_b|^2 \psi_d ~, \\
i \hbar \partial_t \psi_b &=& - \frac{\hbar^2}{2m_b} \nabla^2 \psi_b + g_{bb} |\psi_b|^2 \psi_b + \tilde{g} |\psi_d|^2 \psi_b ~, \label{eq.2gpe}\\ \nonumber
\end{eqnarray}
where the mass of the dimer is $m_d = 2 m_f$ and the coupling constants of the dimer-dimer and dimer-boson interactions, in the first Born approximation, are given by $g_{dd} = 2 g_{ff}$ and $\tilde{g} = 2 g_{bf}$ (note that, though the exact many-body values for these parameters are different~\cite{Giorgini2008}, we use the mean-field expressions for consistency with the BdG equations). Replacing the BdG equation for the order parameter $\Delta$ of paired fermions with a GP equation for the order parameter $\psi_d$ of bosonic dimers is expected to be a good approximation for $1 / (k_F a_{ff})$ of the order of, or larger than one. An important consequence is that the region of the phase diagram of the mixture where one finds the intermediate, partially mixed phase becomes narrower when moving from unitarity towards the BEC regime \cite{Salasnich2007} and eventually disappears in the deep BEC limit. An example is given in Fig.~\ref{fig.bec}, where we show the density profiles of fermions (blue lines) and bosons (red lines) of the Fermi-Bose mixture, obtained by solving the coupled BdG and GP equations~(\ref{eq.bdg})-(\ref{eq.gpe}) for $1 / {k_F a_{ff}} = 1$, the other parameters remaining the same as in the central row of Fig.~\ref{fig.unitarity}. The solution in the top panel corresponds to the case where the phase of the order parameter $\Delta$ is constant, while the one in the bottom panel is obtained by imposing a $\pi$ phase difference and a node at the centre. Two main comments are in order: i) the comparison with Fig.~\ref{fig.unitarity} shows that the partially mixed phase, which was present at unitarity for the same value of $g_{bf}^2 n_0 / (g_{bb} E_F)$, is lost in the BEC regime in favour of a fully separated phase; ii) if we calculate the density profile by solving the coupled GP equations~(\ref{eq.2gpe}) (dashed lines) instead of the the coupled BdG and GP equations~(\ref{eq.bdg})-(\ref{eq.gpe}) (solid lines), we find very similar results. 

These considerations suggest that the unitary regime is the most suitable and interesting for the investigation of dark-bright solitons in Fermi-Bose mixtures. 


\section{Conclusions}
\label{sec.conclusions}

In this work we theoretically study a mixture of Bose and Fermi superfluids by using a mean-field theory based on the solutions of coupled BdG and GP equations. We tune the interaction between fermions in the BCS-BEC crossover, and we also vary the interspecies and intraspecies interactions in order to explore the three regimes of fully mixed, partially mixed and fully separated phases. The focus of the work is on solitonic solutions, and we find that, in the regime of partial mixing, a dark soliton in the Fermi component becomes wider and deeper, taking all bosons in, but maintaining phase coherence. This dark-bright solitonic structure can serve to stabilize dark solitons in Fermi superfuids against snaking instability, but it is also interesting in itself as an example of many-body state where nonlinearity and coherence play a relevant role. Here we have found that bosons favour and amplify an inhomogeneous (soliton-like) order parameter of fermions. A similar mechanism might also favour the realization of the Fulde-Ferrell-Larkin-Ovchinnikov (FFLO) phase in the unitary regime of the Fermi gas, which involves density modulations and could be enhanced by the presence of bosons with spin-dependent interaction, thus making the observation of such elusive phase easier.

We finally notice that the range of parameters in which we find partial mixing may be experimentally accessible using, for instance, mixtures of $^6$Li-$^7$Li, $^6$Li-$^{87}$Rb or $^{23}$Na-$^{87}$Rb. When the mass of the two components is different, and particularly when $m_f / m_b \ll 1$, the situation is expected to be even more favourable. In fact, by expressing Eq.~(\ref{eq.lda-bertsch}) in terms of masses and scattering lengths, one finds that the condition for the stability of the miscible phase becomes $(n_f^{1/3})_{\rm crit} \propto a_{bb}/a_{bf}^2 (m_f/m_b) / [1+ (m_f/m_b)]^2$ and phase separation can be reached for smaller densities, possibly accessible to current experiments.

\begin{acknowledgements}
M.T. thanks Stefano Giorgini and Krzysztof Sacha for useful discussions.
This work is supported by ERC through the QGBE grant, by the QUIC grant of the Horizon2020
FET program and by Provincia Autonoma di Trento. 
A.R. acknowledges support from the Alexander von Humboldt foundation. 
This work is also supported by the PL-Grid infrastructure (M.T.). 
\end{acknowledgements}

\end{document}